\documentclass[prl,aps,superscriptaddress,floatfix,nofootinbib,twocolumn]{revtex4-1}
\pdfoutput=1
\usepackage{exscale}                  
\usepackage[intlimits]{amsmath}       
\usepackage{amsfonts}
\usepackage{amssymb,amscd}
\usepackage[dvips]{epsfig}                   
\usepackage{array}
\usepackage{wrapfig}
\usepackage{color}
\usepackage{subfigure}

\usepackage{graphicx}
\usepackage{rotating}
\usepackage{bm}
\usepackage{dsfont}
\usepackage{slashed}
\usepackage{color}

\newcommand{\beqa}{\begin{eqnarray}}
\newcommand{\eeqa}{\end{eqnarray}}
\newcommand{\beq}{\begin{equation}}
\newcommand{\eeq}{\end{equation}}

\def\eq#1{(\ref{#1})}

\def\Fig#1{Fig.~\ref{#1}}

\def\s0#1#2{\mbox{\small{$ \frac{#1}{#2} $}}}
\def\0#1#2{\frac{#1}{#2}}

\def\CP{{\mathcal P}}

\def\CV{{\mathcal V}}

\newcommand{\tr}{\mathrm{tr}}
\newcommand{\I}{\mathrm{i}}
\newcommand{\be}{\begin{eqnarray}}
\newcommand{\ee}{\end{eqnarray}}

\newcommand{\Nc}{N_{\rm{c}}}

\definecolor{darkgreen}{rgb}{0,0.6,0}
\definecolor{gray}{rgb}{.7,.7,.7}

\begin{document}
\title{Polyakov loop potential at finite density}

\author{Christian~S.~Fischer} \affiliation{Institut f\"ur Theoretische
  Physik, JLU Gie\ss{}en, Heinrich-Buff-Ring
  16, 35392 Gie\ss{}en, Germany} \author{Leonard~Fister}
\affiliation{Department of Mathematical Physics, National University
  of Ireland Maynooth, Maynooth, County Kildare, Ireland} \author{Jan
  Luecker} \affiliation{Institut f\"ur Theoretische Physik,
  JLU Gie\ss{}en, Heinrich-Buff-Ring 16,
  35392 Gie\ss{}en, Germany} \author{Jan~M.~Pawlowski}
\affiliation{Institut f\"{u}r Theoretische Physik, Universit\"{a}t
  Heidelberg, Philosophenweg 16, 69120 Heidelberg, Germany}
\affiliation{ExtreMe Matter Institute EMMI, GSI Helmholtzzentrum
  f\"{u}r Schwerionenforschung mbH, 
  64291 Darmstadt,
  Germany} 

\begin{abstract}
The Polyakov loop potential serves to distinguish between the confined 
hadronic and the deconfined quark-gluon plasma phases of QCD. For
$N_f=2+1$ quark flavors with physical masses we determine the 
Polyakov loop potential at finite temperature and density and extract 
the location of the deconfinement transition. We find a cross-over
at small values of the chemical potential running into a critical
end-point at $\mu/T > 1$.
\end{abstract}

\maketitle

\noindent {\bf Introduction} 

In recent years much progress has been made in our understanding of
the phase structure of QCD at finite temperature and density. This
understanding has been achieved with a variety of methods ranging from
first principle lattice and continuum computations to elaborate model
studies. 

At vanishing density all these methods by now converge quantitatively
leaving only a few open fundamental questions, e.g.\ the order of the
phase transitions in different regions of the Columbia plot. In turn,
at finite density, progress has been hampered by several intricate
problems.  On the lattice one has to face the sign problem which so
far has made it impossible to access chemical potentials with $\mu/T
>1$ \cite{Endrodi:2011gv,Kaczmarek:2011zz}.  First principle continuum
computations with functional methods, such as Dyson-Schwinger
equations (DSE) and functional renormalisation group (FRG) equations,
are based on an expansion of the theory in terms of quark-gluon
correlation functions. Hence at finite density they have to cope with
the increasingly complicated ground state structure of QCD in terms of
these correlation functions. Finally, low energy effective models are
usually anchored and benchmarked at the vacuum and thermal physics at
vanishing density. In turn, the more important the density
fluctuations get, the less quantitative are the results.

Facing these problems, it is apparent that progress in our
understanding of QCD at finite temperature and density is probably
best achieved by a combination of the different methods at hand. In
the present work we push forward the functional continuum approach
towards the phase diagram of QCD supplemented with results from
lattice QCD. We determine, for the first time, the Polyakov loop
potential at finite temperature and real chemical potential.\\[-2ex]

\noindent {\bf The phase diagram with functional methods}

In the past decade continuum quark and gluon correlations functions
have been computed with the help of functional equations for the
effective action of QCD. These works have been mostly performed in
(background) Landau gauge,
\begin{equation}\label{eq:backlandau} 
  \bar D_\mu A_\mu =0\,,\quad {\rm with}\quad \bar D_\mu 
  =\partial_\mu -i g \bar A_\mu\,, 
\end{equation}
where $\bar A$ is chosen to be the expectation value of the gauge
field, $\bar A=\langle A\rangle$. The present work also utilizes the
gauge \eq{eq:backlandau}. Correlation functions in ordinary Landau
gauge are directly related to those in background Landau gauge by
simply substituting plain momentum $p^2$ with background covariant
momentum, $p^2\to - \bar D^2$ \cite{Braun:2007bx,Fister:2013bh}.  In
this approach the Polyakov loop variable 
\beq L=\frac{1}{\Nc} \tr_{\rm fund}\, P(\vec x)\,,\quad {\rm with}\quad 
P(\vec x)= \CP\, e^{\I g \int_0^\beta dx_0\,
    {A}_0(x_0,\vec{x})}\,,
\label{eq:Polloop}
\eeq 
in the fundamental representation, evaluated at the minimum of the
Polyakov loop effective potential $V[A_0]$, is an order parameter for confinement 
\cite{Braun:2007bx,Marhauser:2008fz}. 
The effective potential is defined from the effective action $\Gamma$, evaluated 
at constant background fields $A_0^{\rm const}$ and vanishing gauge fields, 
\begin{equation}\label{eq:Vpol}
V[A_0^{\rm const}]:=\0{1}{\beta \rm \CV} \Gamma[A^{\rm const}_0;0]\,.
\end{equation} 
The minimum of $V[A_0]$ singles out the expectation
value of the gauge field in the background Landau gauge, $\langle
A_0\rangle$. The related order parameter satisfies
\beq L[\langle A_0\rangle]\geq  \langle L[A_0]\rangle 
\label{eq:PolloopJensen}
\eeq 
within an appropriate (re)normalization of $ \langle L[A_0]\rangle$,
see \cite{Braun:2007bx,Marhauser:2008fz,Braun:2010cy,Fister:2013bh}. This
inequality holds true for both, Yang-Mills theory and fully dynamical
QCD. In the presence of a phase transition both sides vanish at $T_c$
and the inequality \eq{eq:PolloopJensen} is saturated below $T_c$. In turn, in
the presence of a cross-over we expect the cross-over temperature
computed from $L[\langle A_0\rangle]$ to be lower than the one
computed from $\langle L[A_0]\rangle$. 

\begin{figure}[t]
\begin{center}
\includegraphics[width=.95\columnwidth]{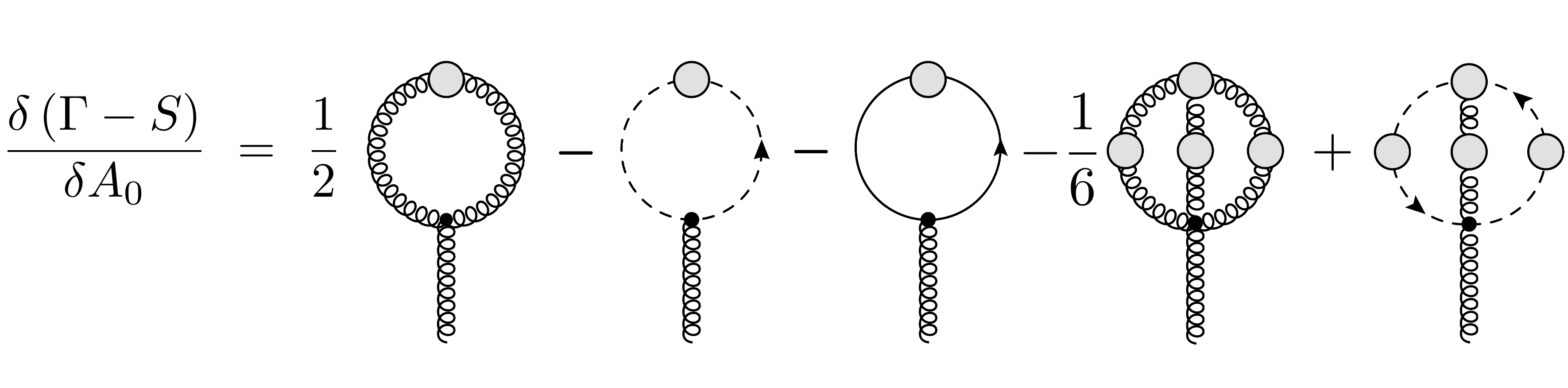}
\caption{DSE for the background gluon one-point function. Large circles
indicate dressed propagators and vertices, and $S$ stands for the classical action, 
see \cite{Fister:2013bh}.}
\label{fig:DSE-A}
\includegraphics[width=.85\columnwidth]{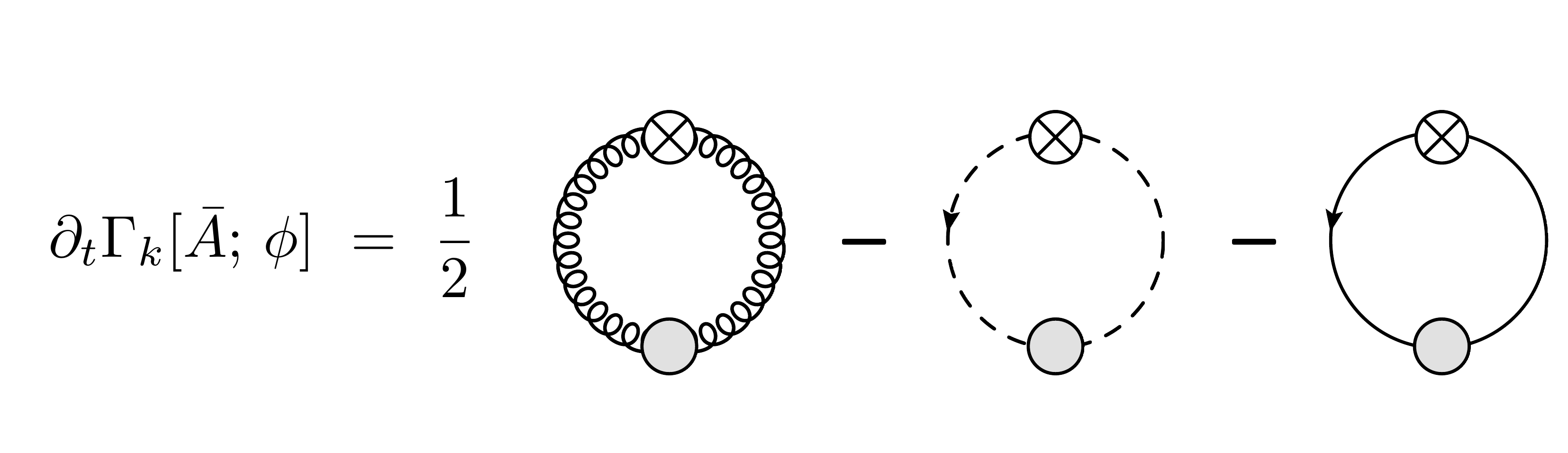}
\caption{Functional flow for the effective action of QCD. Crosses
  indicate insertions of the functional cut-off. The field
  $\phi$ combines quark, ghost and gluon fields.}\vspace*{-5mm}
\label{fig:FRG-QCD}
\end{center}
\end{figure}
The effective potential, or its $A_0$-derivative, can be computed from
the functional DSE and FRG equations, see \Fig{fig:DSE-A} and
\Fig{fig:FRG-QCD}, respectively.  For the FRG this has been put
forward in Yang-Mills theory,
\cite{Braun:2007bx,Marhauser:2008fz,Braun:2010cy,Fister:2013bh}, and
in QCD at finite temperature and imaginary chemical potential in
\cite{Braun:2009gm}. There, the effective potential $V[A_0]$ is
computed solely from the scale-dependent propagators. More recently, a
similar computation of the Polyakov loop potential has also been
performed in Coulomb gauge, \cite{Reinhardt:2012qe,
  Reinhardt:2013iia}. Related lattice computations can be found 
in \cite{Langelage:2010yr,Diakonov:2012dx,Greensite:2012dy,Greensite:2013yd}. 

In turn, the DSE-formulation has been put forward in
\cite{Fister:2013bh}. It is apparent from \Fig{fig:DSE-A}, that the
effective potential $V[A_0]$ can be computed from the DSE once the
ghost, gluon and quark propagators as well as the three-gluon
vertex and ghost-gluon vertex are known. In the present work we
utilize the observation in Ref.~\cite{Fister:2013bh} that within an
optimized renormalisation scheme the two-loop terms in \Fig{fig:DSE-A}
are sub-leading at temperatures about $T_c$.  This has been thoroughly
tested for Yang-Mills theory within a comparison of the DSE results
from \Fig{fig:DSE-A} with the FRG results from \Fig{fig:FRG-QCD}. For
temperatures about $T_c$ the results agree quantitatively. The
inclusion of the quark-loop in full QCD does not change this picture.
Moreover, we neglect the $A_0$-dependence of the back-reaction of the
Polyakov loop potential to the chromo-electric propagator in terms of
$\partial_{A_0}^2 V[A_0]$. While these back-reaction effects may be
crucial for the critical scaling of the chromo-electric component of
the gluon propagator close to the phase transition of pure Yang-Mills
theory \cite{Maas:2011ez,Fister:2013bh}, we expect its influence on
the QCD transition to be small. This needs to be verified in future
work.

Within this approximation the $A_0$-dependence
solely originates from the shifted Matsubara frequencies $p_0+ g
A_0$. The diagrams in \Fig{fig:DSE-A} and \Fig{fig:FRG-QCD} can be
diagonalized in color space leaving us with
\begin{equation}\label{eq:matsu}
p_0 +2 \pi T  \varphi_m \,,
\end{equation} 
where $\varphi_m$ are the eigenvalues of $\beta g A_0/(2 \pi)$,
depending on the representation. For example, for two-color
QCD, the constant temporal gauge
field can be rotated into the Cartan subalgebra, $A_0= 2 \pi
T\varphi/g\, \tau^3$, with Cartan generator $\tau^3$.  We have the
eigenvalues
\begin{equation}\label{eq:evsu2}
  \varphi_{\rm ad}\in \{\pm \varphi,0\}\,,\qquad  
  \varphi_{\rm fund}\in \left\{\pm \0{\varphi}{2} \right\}\,,
\end{equation}
in the adjoint and fundamental representation respectively. The
factors $1/2$ in $\varphi_{\rm fund}$ carry information
on the explicit center-symmetry breaking of the quarks.

In the physical case of $SU(3)$ we restrict ourselves to $2 \pi
T\varphi/g\, \tau^3$ in the Cartan subalgebra generated by
$\tau^3,\tau^8$.\footnote{At finite chemical potential, this involves
  a center average.} The corresponding eigenvalues are given by
\begin{equation}\label{eq:evsu3}
  \varphi_{\rm ad}\in \left\{\pm\varphi ,\pm\0{\varphi}{2},\pm
    \0{\varphi}{2}, 0,0\right\}\,,\qquad  
  \varphi_{\rm fund}\in \left\{ \pm\0{\varphi}{2}, 0\right\}\,,
\end{equation}
for more details see \cite{Fister:2013bh,Braun:2010cy}. Then, the
shifted Matsubara frequencies $p_0+g A_0$ read after diagonalization,
\begin{equation} \label{eq:matsubaraSU3} 2\pi T(n + \varphi_{\rm
    ad})\,,\qquad {\rm and}\qquad 2\pi T\left(n + \012+ \varphi_{\rm
      fund}\right)\,,
\end{equation} 
for ghost, gluon in the adjoint representation and the quark in the
fundamental representation respectively. The additive nature of the
loop representation in \Fig{fig:DSE-A} and \Fig{fig:FRG-QCD} leads to
the simple form
\begin{equation}\label{eq:fullpot} 
V(\varphi)=V_{\rm glue}(\varphi) + V_{\rm quark}(\varphi)\,. 
\end{equation}
Here, $V_{\rm glue}$, contains all contributions
from the gluon and ghost diagrams in the DSE and FRG, see Figs.\ref{fig:DSE-A},\ref{fig:FRG-QCD}. 

In the present approximation, i.e. without the
backreaction of $V[A_0]$ to the chromo-electric gluon, all diagrams
contributing to $V_{\rm glue}$ involve only traces and contractions in
the adjoint representation, and hence the eigenvalues $\varphi_{\rm
  ad}$ in \eq{eq:evsu2}, \eq{eq:evsu3}. In turn, the matter
contribution, $V_{\rm quark}$, involves only traces and contractions
in the fundamental representation, and hence the eigenvalues
$\varphi_{\rm fund}$ in \eq{eq:evsu2}, \eq{eq:evsu3}. With
\eq{eq:matsubaraSU3} this leads to the periodicities
\begin{equation}\label{eq:potperiodicsu3} 
V_{\rm glue}(\varphi+2)=V_{\rm glue}(\varphi)\,,\quad\quad  
V_{\rm quark}(\varphi+2)= V_{\rm quark}(\varphi)\,,  
\end{equation}
for the physical case of $SU(3)$. For comparison we also quote the
$SU(2)$-case where we have
\begin{equation}\label{eq:potperiodicsu2} 
V_{\rm glue}(\varphi+1)=V_{\rm glue}(\varphi)\,,\quad\quad  
V_{\rm quark}(\varphi+2)= V_{\rm quark}(\varphi)\,. 
\end{equation}
We observe that the periodicity of $V_{\rm quark}$ is independent of
$N_c$ in contrast to that of the glue part. The latter dependence
reflects the fact that $V_{\rm glue}$ is center-symmetric and hence
invariant under $Z_{N_c}$-transformations.  For the simple case of
$N_c=2$ the Cartan is one-dimensional and a center transformation   
entails $\varphi \to 1- \varphi$ with center-symmetric point
$\varphi=1/2$. Evidently this is not the symmetry of the quark
potential $V_{\rm quark}$ due to its periodicity, see
\eq{eq:potperiodicsu2}. The Polyakov loop in the fundamental
representation in $SU(2)$ reads
\begin{equation}\label{eq:polloopvarphisu2}
L(\varphi)=\cos (\pi\varphi)\,, 
\end{equation}
and vanishes at the center-symmetric point $\varphi=1/2$.

For $N_c=3$ (and higher $N_c$) a center transformation is a rotation
in the Cartan. Accordingly, the explicit center-breaking in the quark
potential is only visible for general gauge fields in the Cartan
sub-algebra, {\it i.e.} $A_0=A_0^3\tau^3+A_0^8\tau^8$, which are not
considered here. Interestingly, for $SU(3)$ the quark potential has
the same periodicity w.r.t.\ $\varphi$ as the glue potential in
contradistinction to $SU(2)$. This may be a helpful property for model
applications at finite density, \cite{Pisarski:2000eq, Dumitru:2003hp,
  Herbst:2010rf,Skokov:2010uh,Weise:2012yv, Bratovic:2012qs,
  Fukushima:2012qa,Haas:2013qwp}, and shall be studied elsewhere.
The Polyakov loop in three-color QCD reads
\begin{equation}\label{eq:polloopvarphi}
L(\varphi)=\frac{1}{3}\left(1+2 \cos (\pi\varphi)\right)\,, 
\end{equation}
and vanishes at the confining values $\varphi=2/3,4/3$ in the fundamental period 
$\varphi\in\{0,2\}$. This gives us direct access to an order
parameter potential for the confinement-deconfinement phase transition
in a DSE-approach to the phase structure of QCD as put forward in
\cite{Fischer:2011mz,Fischer:2012vc}.  In the following we will
exploit this approach at finite temperature and density thus providing
first insights into the Polyakov loop potential at finite density.\\[-2ex]

\noindent {\bf DSE for the quark and gluon propagators}

\begin{figure}[t]
\centering\includegraphics[width=0.4\textwidth]{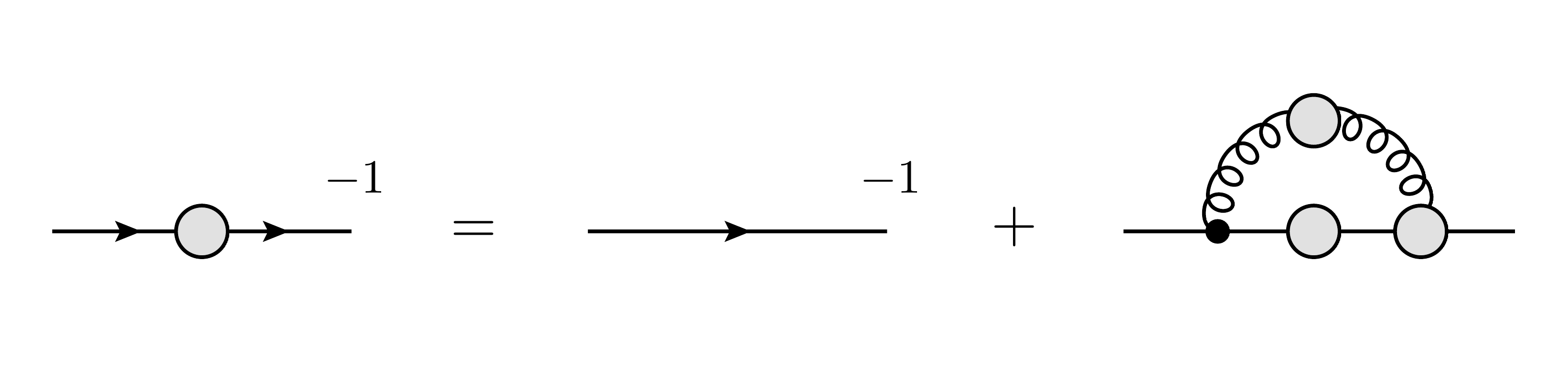}
\caption{The DSE for the quark propagator.}
\label{fig:quarkDSE}
\includegraphics[width=0.4\textwidth]{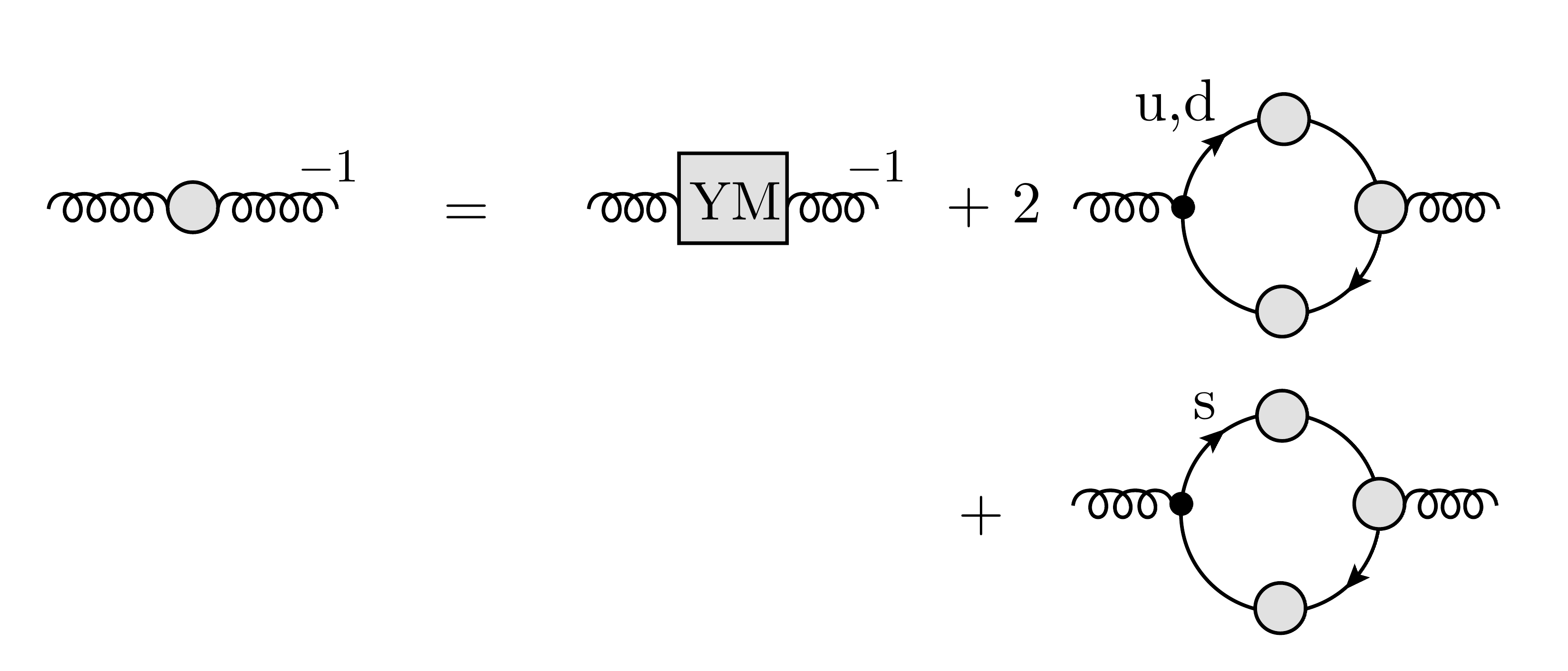}
\caption{The truncated gluon DSE for $N_f=2+1$ QCD. The first term is the 
inverse quenched propagator.} 
\label{fig:apprGluonDSE}
\end{figure}

In order to determine the $N_f=2+1$ quark and gluon propagators at finite 
temperature and chemical potential we have solved their corresponding 
DSEs given diagrammatically in Figs.~\ref{fig:quarkDSE} 
and \ref{fig:apprGluonDSE}.
\begin{figure}[t]
\includegraphics[width=0.40\textwidth]{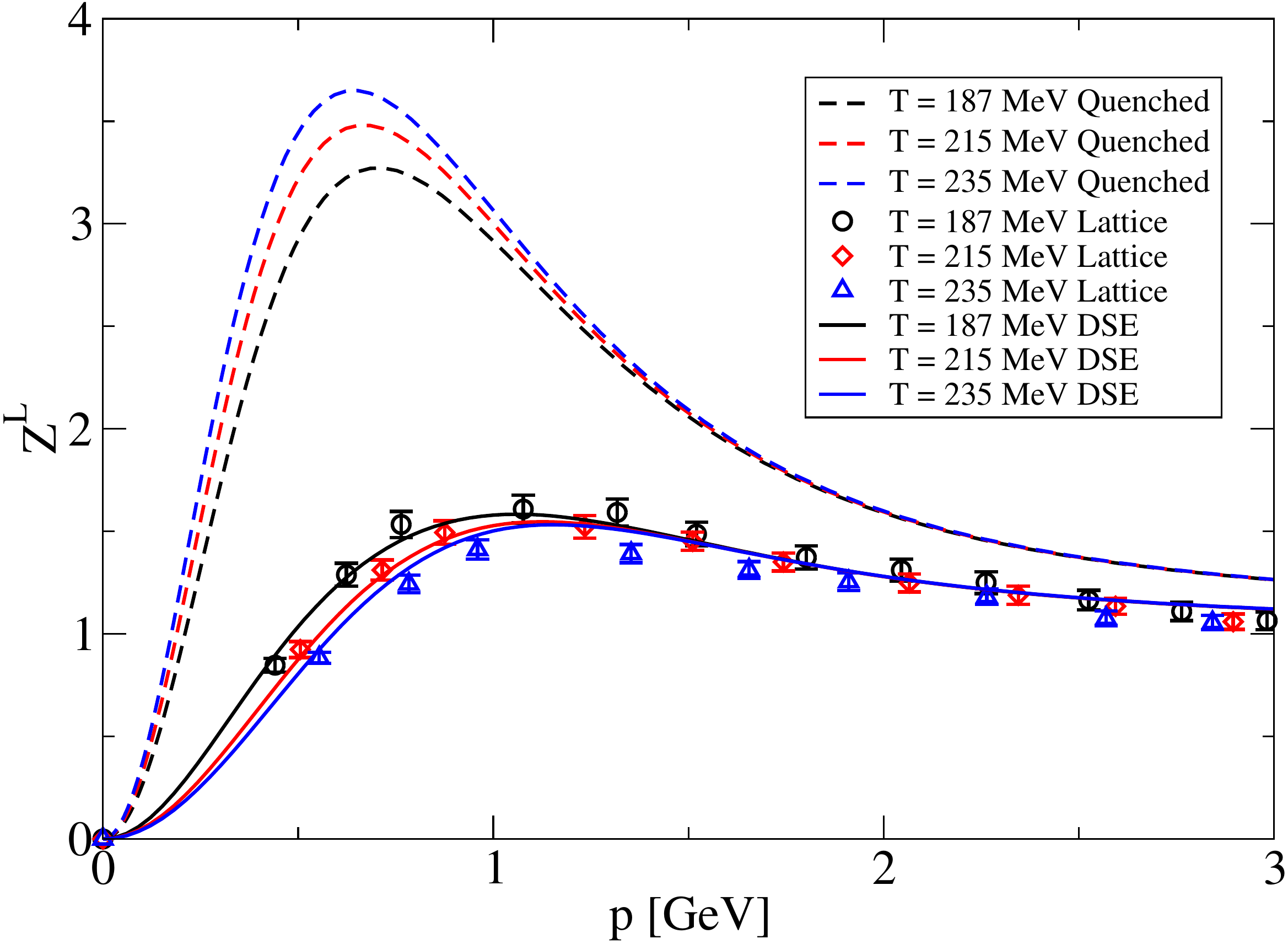}
\includegraphics[width=0.40\textwidth]{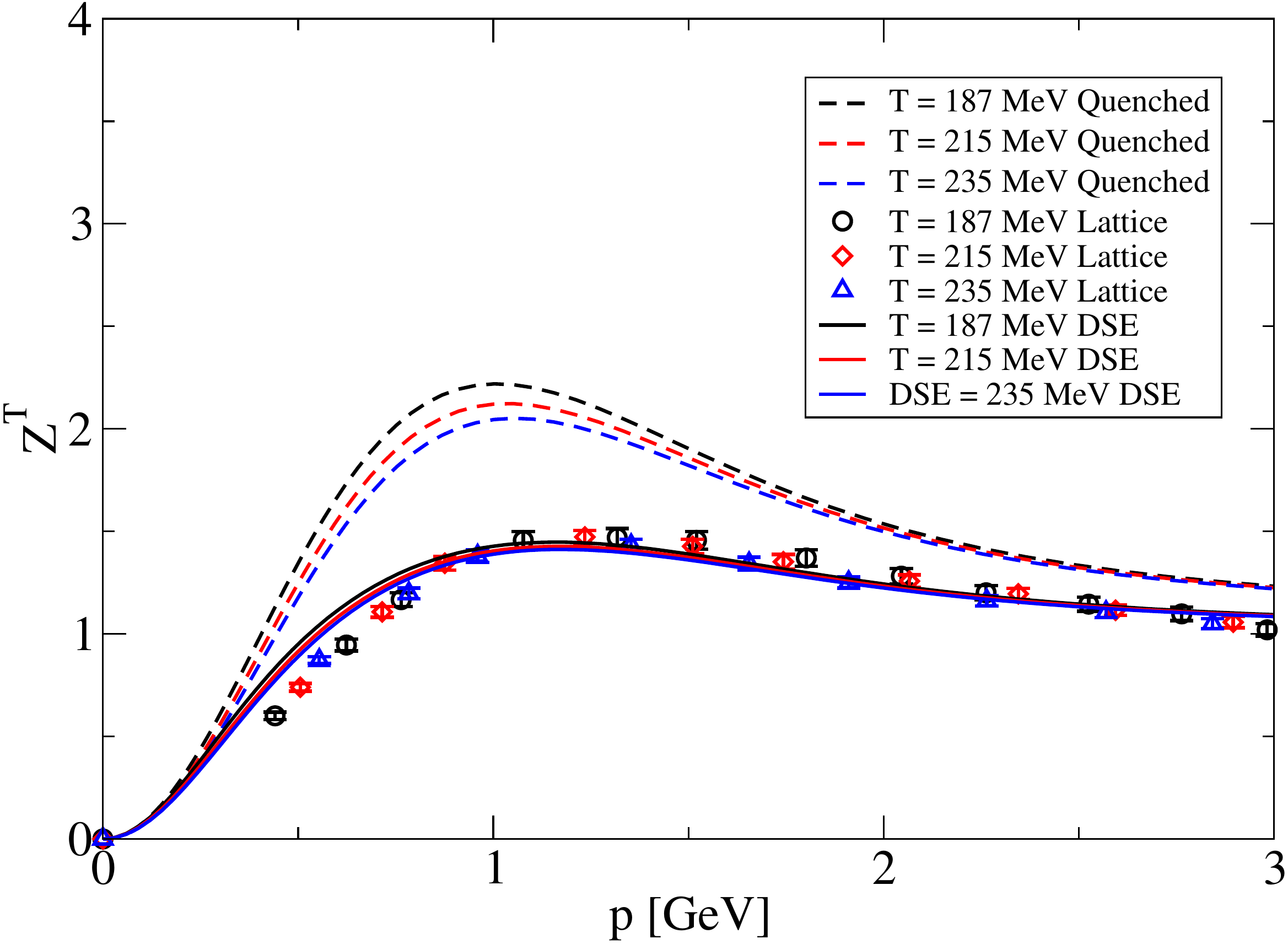}
\caption{Quenched and unquenched gluon dressing functions $Z^L$ (upper
  plot) and $Z^T$ (lower plot), see \eq{eq:gluonprop}, compared to
  gauge-fixed unquenched lattice data from
  \cite{Aouane:2012bk}. }
	\label{fig:glue}
\end{figure}
In the gluon DSE we work with an approximation neglecting unquenching
effects in the Yang-Mills part of the equation. Consequently this part
can be replaced by the inverse quenched propagator denoted by the
diagram with the box labelled 'YM' in
Fig.~\ref{fig:apprGluonDSE}. This approximation is valid on the few
percent level \cite{Fischer:2012vc}. For the quenched gluon propagator
one may use corresponding lattice results
\cite{Fischer:2010fx,Aouane:2011fv,Maas:2011ez,Cucchieri:2012nx} or
input from a FRG calculation within Yang-Mills theory
\cite{Fister:2011uw, Fister:2013bh}. We have checked that our results
for the potential and the respective critical temperatures are hardly
affected by this choice. This is a direct consequence of the
inheritance of the above-mentioned renormalisation scheme in the
quenched case \cite{Fister:2013bh}, allowed by the absence of two-loop
diagrams in the matter sector of the DSE.  To make contact with the
results of Ref.~\cite{Fischer:2012vc} in the following we use the
lattice results of Ref.~\cite{Fischer:2010fx} as input. The only other
unknown quantity in our system of DSEs is the fully dressed
quark-gluon vertex. Since no reliable calculations of this quantity at
finite temperature are available, we resort to the model ansatz of
Refs.~\cite{Fischer:2011mz,Fischer:2012vc}. There, the vertex is
constructed utilizing information from the Slavnov-Taylor identity of
the vertex as well as constraints due to the perturbative RG running
of the vertex. It has been shown in \cite{Fischer:2012vc} that such an
ansatz is sufficient to deliver results for the chiral condensate at
finite temperature in good agreement with lattice gauge theory
\cite{Borsanyi:2010bp}.  A further justification of our quark-gluon
interaction is given in Fig.~\ref{fig:glue}.  In the thermal medium,
the color-diagonal gluon propagator is given by
\begin{eqnarray}\label{eq:gluonprop}
D_{\mu\nu}(p) &=& P_{\mu\nu}^L(p)\frac{Z^L(p)}{p^2} + P_{\mu\nu}^T(p)\frac{Z^T(p)}{p^2}\,,
\end{eqnarray}
where the dressing functions $Z^L$ and $Z^T$ represent the parts with longitudinal 
and transversal orientation with respect to the heat bath and the $P_{\mu\nu}^{T,L}$
are the corresponding projectors. For three different temperatures these dressing
functions are plotted in Fig.~\ref{fig:glue}. The dashed lines are fits to the quenched
lattice data of \cite{Fischer:2010fx}. 
The unquenched results (solid lines), predicted in the DSE framework \cite{Fischer:2012vc},
are compared with very recent unquenched lattice results from Ref.~\cite{Aouane:2012bk}. 
We observe large unquenching effects in the longitudinal part of the propagator
and somewhat smaller effects in the magnetic part. For both dressing functions
the prediction from the functional framework is nicely matched by the lattice data.
We believe these results provide solid justification for the vertex construction
and the truncation of the gluon DSE used in our work. \\[-2ex]
%
\begin{figure}[t]
\includegraphics[width=0.38\textwidth]{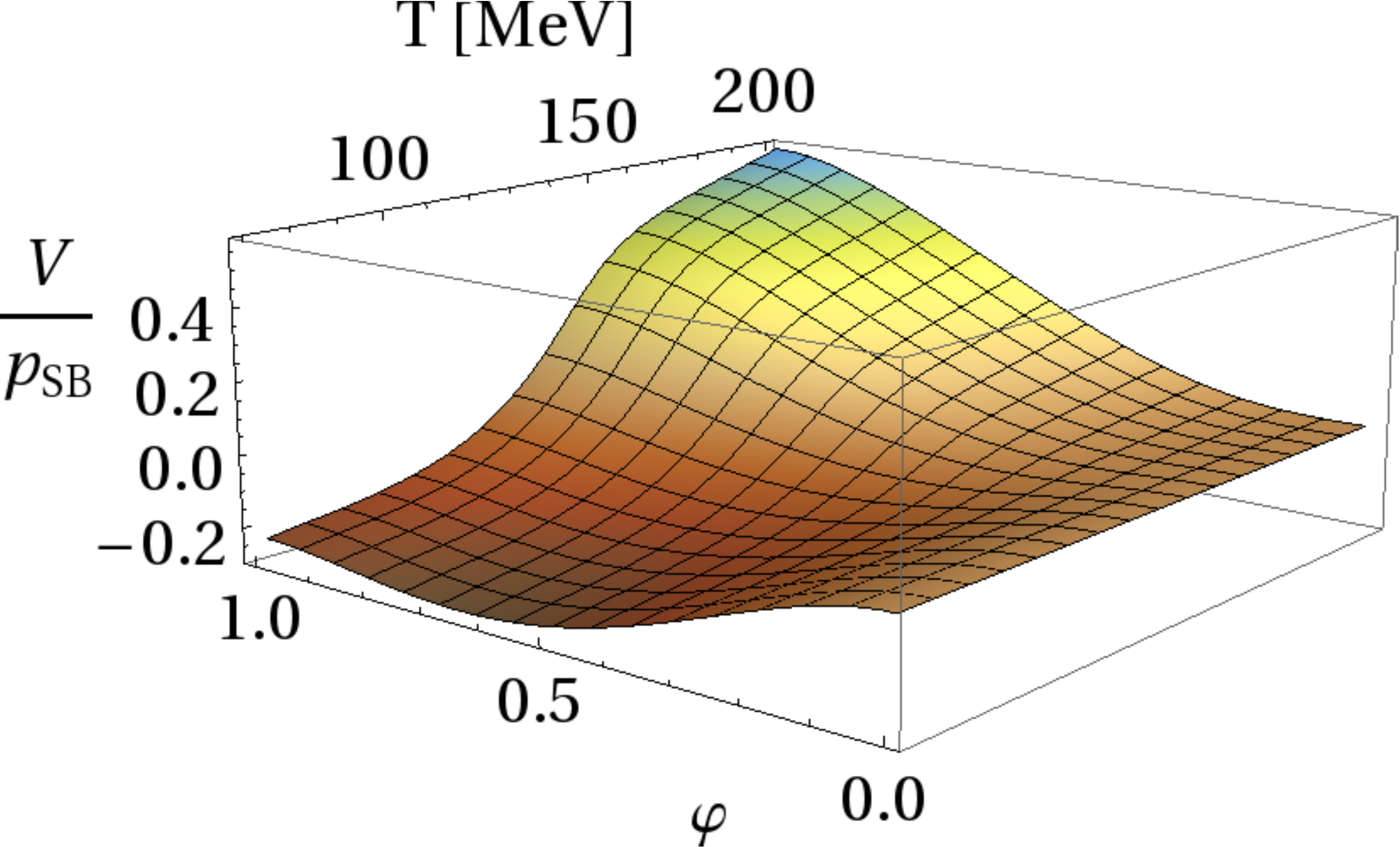}
\caption{Polyakov loop potential defined in \eq{eq:Vpol} for $\mu=0$.}
\label{fig:PotentialMu0}\end{figure}
\begin{figure}[t]
\includegraphics[width=0.38\textwidth]{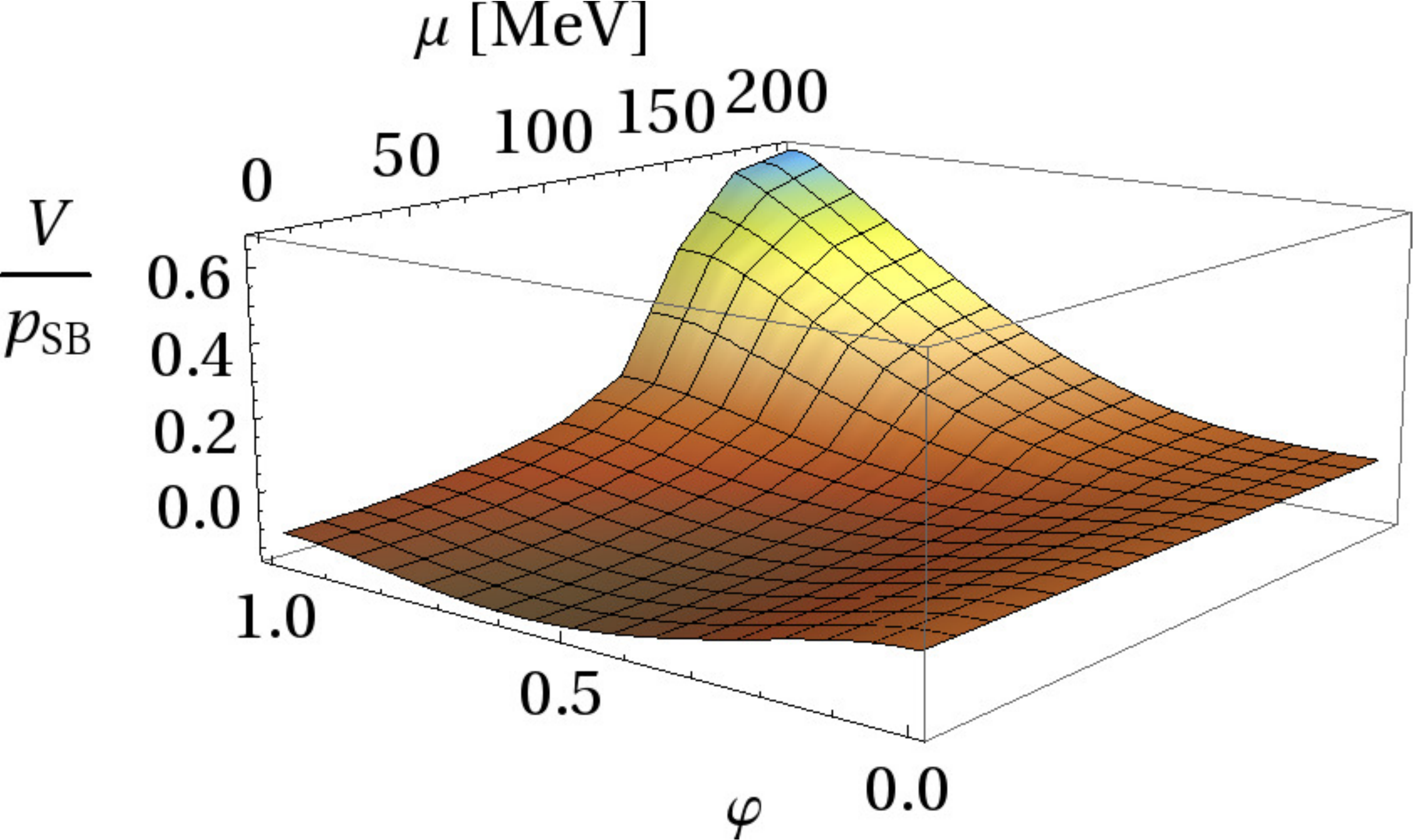}
\caption{Polyakov loop potential for $T=115$ MeV.}
\label{fig:PotentialT115}\end{figure}
%

\noindent{\bf Results} 

The DSE for the potential depicted in \Fig{fig:DSE-A} is used to
compute $\partial_\varphi V(\varphi)$. Upon $\varphi$-integration 
this yields the Polyakov loop potential $V(\varphi)$ as a function
of temperature and chemical potential. In \Fig{fig:PotentialMu0} and
\Fig{fig:PotentialT115} we show the dimensionless potential
$V(\varphi)/p_{SB}$ with $V(0)=0$ and $p_{SB}=\frac{19\pi^2}{36}T^4 +
\frac{3}{2}T^2\mu^2 + \frac{3}{4\pi^2}\mu^4$. The pressure is hidden 
in the integration constant \cite{Fister:2013bh} and will be discussed elsewhere.

We have computed the Polyakov loop potential $V(\varphi)$ in 2+1 
flavor QCD at the physical pion mass. The confining
minimum with vanishing Polyakov loop, $L(\varphi)=0$, is at
$\varphi=2/3$, see \eq{eq:polloopvarphi}. In turn, for $\varphi=0$ we
have $L(\varphi=0)=1$. One clearly sees the transition from the
confining regime at low temperature/small chemical potential to the
deconfined phase at high temperature/large chemical potential.  The
sharper cross-over transition as a function of chemical potential with
fixed $T=115$ MeV reflects the proximity of the critical endpoint.
\begin{figure}[t]
\includegraphics[width=0.38\textwidth]{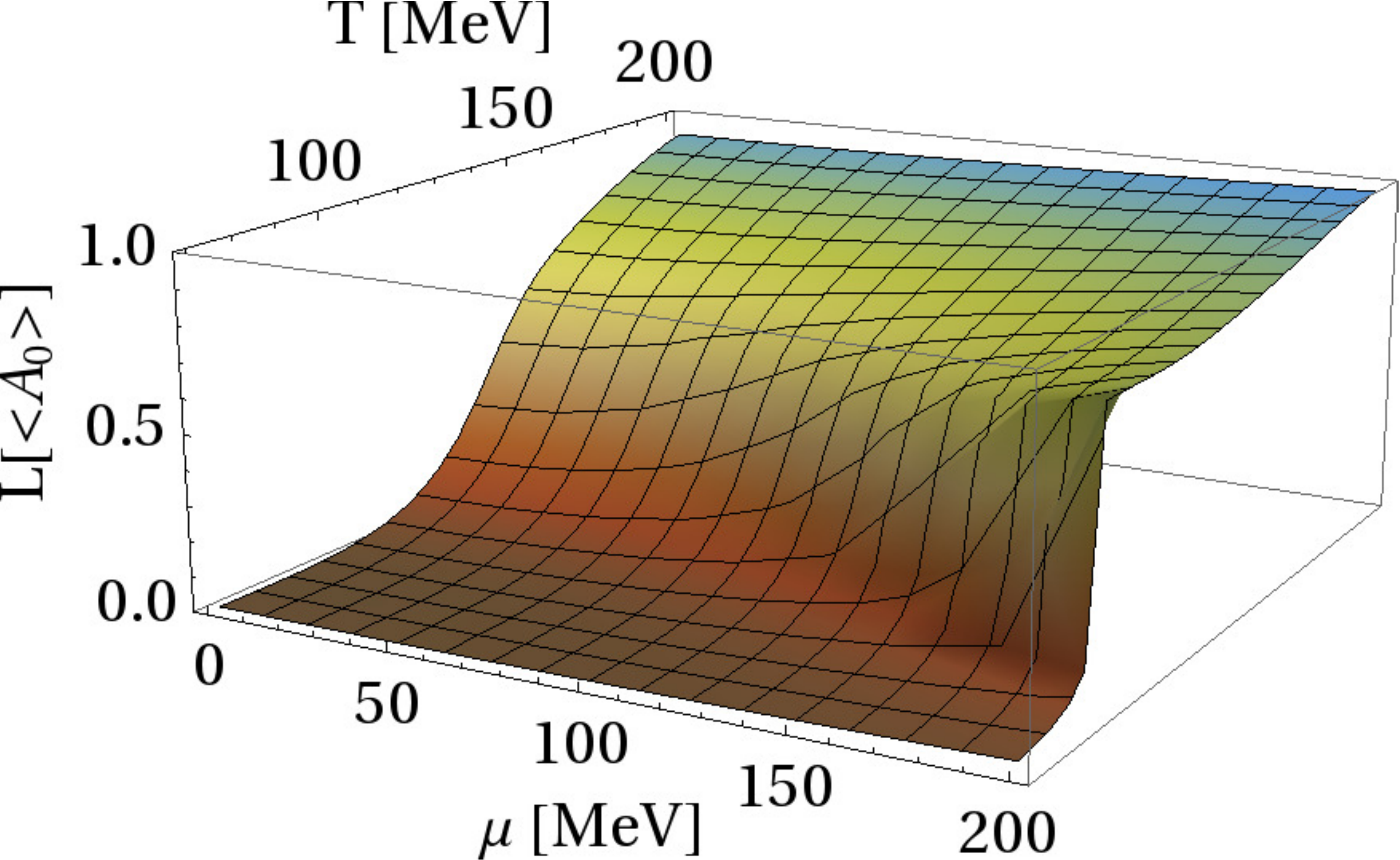}
\caption{Polyakov loop in the $\mu-T$-plane.}
\label{fig:PolL3d}
\end{figure}
%
\begin{figure}[t]
\includegraphics[width=0.38\textwidth]{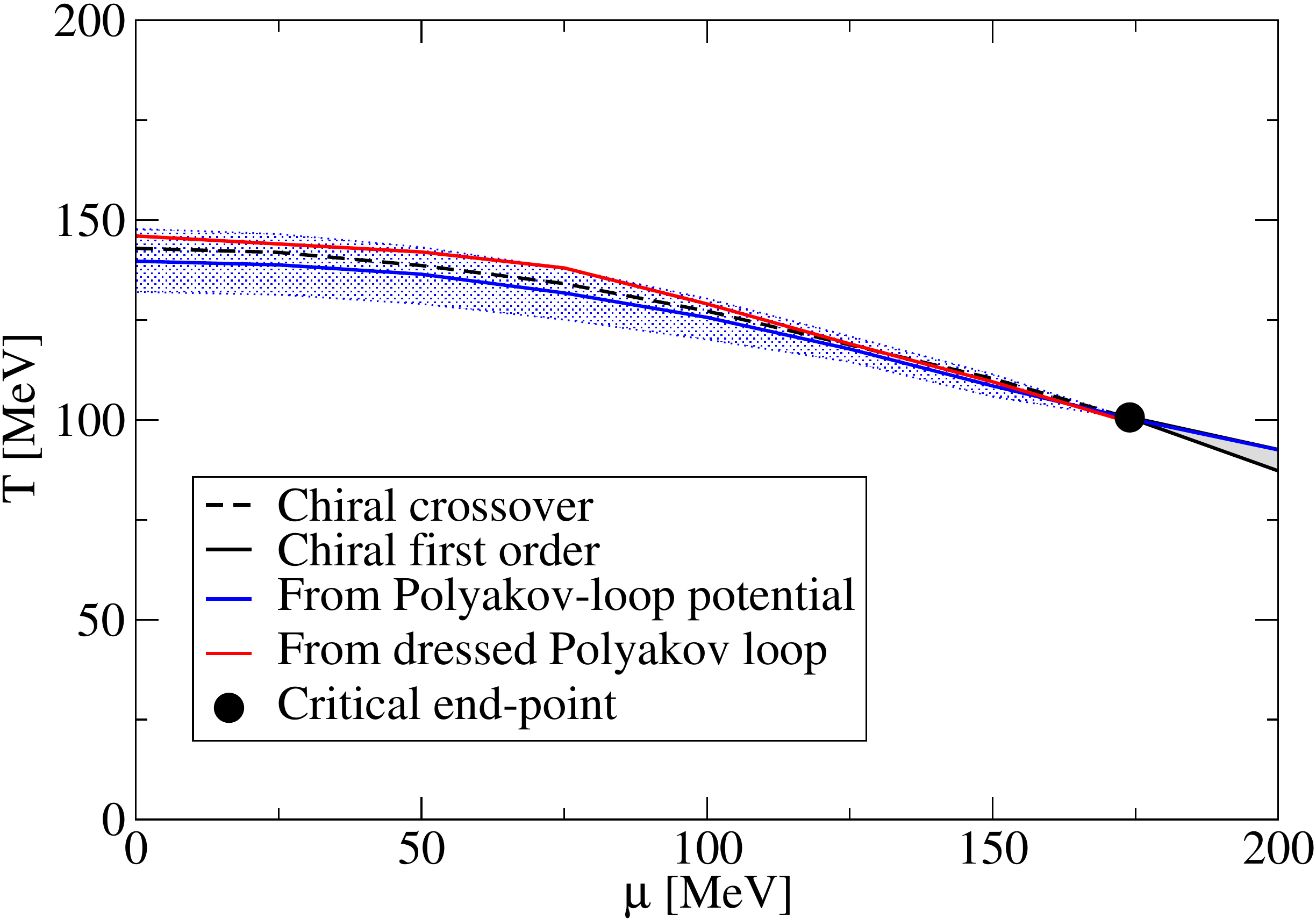}
\caption{Phase diagram for chiral symmetry restoration and
  deconfinement for $N_f=2+1$.}
\label{fig:phasediag}
\end{figure}
%

\Fig{fig:PolL3d} shows the Polyakov loop \eq{eq:Polloop} evaluated at
the minimum $\langle A_0\rangle $ of the effective potential
$V[A_0]$. For small chemical potential or densities the deconfinement
transition is a smooth cross-over. There is no unique definition of
the cross-over temperature $T_{\rm conf}$. In the present work we use
the inflection point of the Polyakov loop,
\begin{equation}\label{eq:inflect}
\left. \partial_T L[\langle A_0\rangle ] \right|_{ T_{\rm conf}} \geq  
\partial_T L[\langle A_0\rangle ] \,.
\end{equation}
i.e., the maximum of the thermal derivative. Other
definitions include the inflection point of the expectation value
$\langle A_0\rangle $, and that of the dual chiral condensate as
computed in \cite{Fischer:2012vc} for 2+1 flavors. In
\cite{Fischer:2012vc} the cross-over temperature is computed from the
susceptibility and differs slightly from the dual $T_{\rm conf}$
computed here. Also the quark masses have been slightly larger
than the physical ones; this has been corrected in the present work.
The cross-over sharpens with increasing chemical potential and
finally turns into a first order transition at $(T_*, \mu_*)=(101
\mbox{ MeV},174 \mbox{ MeV})$. Note that the critical point
$(T_*,\mu_*)$ as well as the first order line does not depend on the
definition of the cross-over temperature. In \Fig{fig:phasediag} we
show $T_{\rm conf}$ together with the chiral transition temperature
$T_\chi$ which is obtained from the inflection point of the
light-quark condensate.  The shaded area shows the width of the
deconfinement cross-over defined by 80\% of the inflection
peak. Interestingly, all transition temperatures, $T_{\rm conf}$ and
$T_\chi$ agree within this width for the whole phase diagram. Since
definitions of $T_{\rm conf}$ with either Polyakov loop potential or
dressed Polyakov loop are based on different properties of the quark
and gluon correlation functions, this provides a highly non-trivial
check of the self-consistency of the present
approximation. Nevertheless, at very large chemical potential the
present scheme may not be sufficient, see Ref.~\cite{Fischer:2012vc}
for a more detailed discussion.

In this work we presented the first results for the Polyakov loop
potential at finite chemical potential in QCD with $N_f=2+1$,
evaluated from a combination of functional and lattice
methods. Besides providing input for model applications, our results
serve as a benchmark prediction for future
evaluations of the potential with different methods.\\[-2ex]

{\bf Acknowledgements} We thank Jens Braun and Bernd-Jochen Schaefer
for discussions.  This work is supported by the Helmholtz Alliance
HA216/EMMI and by ERC-AdG-290623 as well as the Helmholtz
International Center for FAIR within the LOEWE program of the State of
Hesse, and the Helmholtz Young Investigator Group No.~VH-NG-332.  LF
is supported by the Science Foundation Ireland in respect of the
Research Project 11-RFP.1-PHY3193.

\bibliography{polpot.bib}

\end{document}